\documentclass[11pt]{article}
\usepackage{amsmath,amssymb,graphicx,url}
\begin{document}
\newcommand{\PG}{Painlev\'e-Gullstrand}
\newcommand{\be}{\begin{equation}}
\newcommand{\ee}{\end{equation}}
\newcommand{\bea}{\begin{eqnarray}}
\newcommand{\eea}{\end{eqnarray}}
\newcommand{\bd}[2]{{#1}_{\scriptscriptstyle {\rm #2}}}
\newcommand{\bu}[2]{{#1}^{\scriptscriptstyle {\rm #2}}}
\newcommand{\Rd}{R_d}
\newcommand{\udb}{u_{\scriptscriptstyle db}}
\newcommand{\etainf}{\eta_{\,\infty}}
\newcommand{\mun}{m_u}
\newcommand{\MUN}{M_u}
\newcommand{\msun}{m_{\scriptscriptstyle\odot}}
\newcommand{\MSUN}{M_{\odot}}
\newcommand{\Tm}{T_{2+}}
\newcommand{\Tcs}{T_{cs}}
\newcommand{\Tcso}{T_{cs0}}
\newcommand{\thetad}{\theta_d}
\newcommand{\phid}{\phi_d}
\newcommand{\thetac}{\theta_c}
\newcommand{\phic}{\phi_c}
\newcommand{\tE}{t_{\scriptscriptstyle E}}
\newcommand{\tI}{t_{\scriptscriptstyle I}}
\newcommand{\rpond}{\bd{\rho}{pond}}
\newcommand{\rI}{\rho_{\scriptscriptstyle I}}
\newcommand{\rO}{\rho_{\,\scriptscriptstyle 0}}
\newcommand{\RI}{R_{\scriptstyle I}}
\newcommand{\ri}{r_{\scriptscriptstyle I}}

\begin{center}

{\LARGE {\bf Finite cosmology and a CMB cold spot}}

\vspace{1cm}

{\Large Ronald J. Adler,$^{\ast}$ James D. Bjorken$^{\dag}$ and
        James M. Overduin$^{\ast}$}

\vspace{5mm}

{\large $^{\ast}$Gravity Probe B, Hansen Experimental Physics Laboratory,
Stanford University, Stanford, CA 94305, U.S.A.}

{\large $^{\dag}$Stanford Linear Accelerator Center,
Stanford University, Stanford, CA 94309, U.S.A.}

\end{center}

\vspace{1cm}

\begin{quote}
The standard cosmological model posits a spatially flat universe of infinite
extent.  However, no observation, even in principle, could verify that the
matter extends to infinity.  In this work we model the universe as a finite
spherical ball of dust and dark energy, and obtain a lower limit estimate of
its mass and present size: the mass is at least $5\times10^{23}\MSUN$ and the
present radius is at least 50~Gly.  If we are not too far from the dust-ball
edge we might expect to see a cold spot in the cosmic microwave background,
and there might be suppression of the low multipoles in the angular power
spectrum.  Thus the model may be testable, at least in principle.  We also
obtain and discuss the geometry exterior to the dust ball; it is
Schwarzschild-de~Sitter with a naked singularity, and provides an interesting
picture of cosmogenesis.  Finally we briefly sketch how radiation and
inflation eras may be incorporated into the model.
\end{quote}

\vspace{1cm}

\section{Introduction}

The standard or ``concordance'' model of the present universe has been very
successful in that it is consistent with a wide and diverse array of
cosmological data.  The model posits a spatially flat ($k=0$)
Friedmann-Robertson-Walker (FRW) universe of infinite extent, filled with
dark energy, well described by a cosmological constant, and pressureless
cold dark matter or ``dust.''  Despite the phenomenological success of the
model, our present ignorance of the physical nature of both the dark energy
and dark matter should prevent us from being complacent.

\begin{figure}[t!]
\begin{center}
\includegraphics[width=115mm]{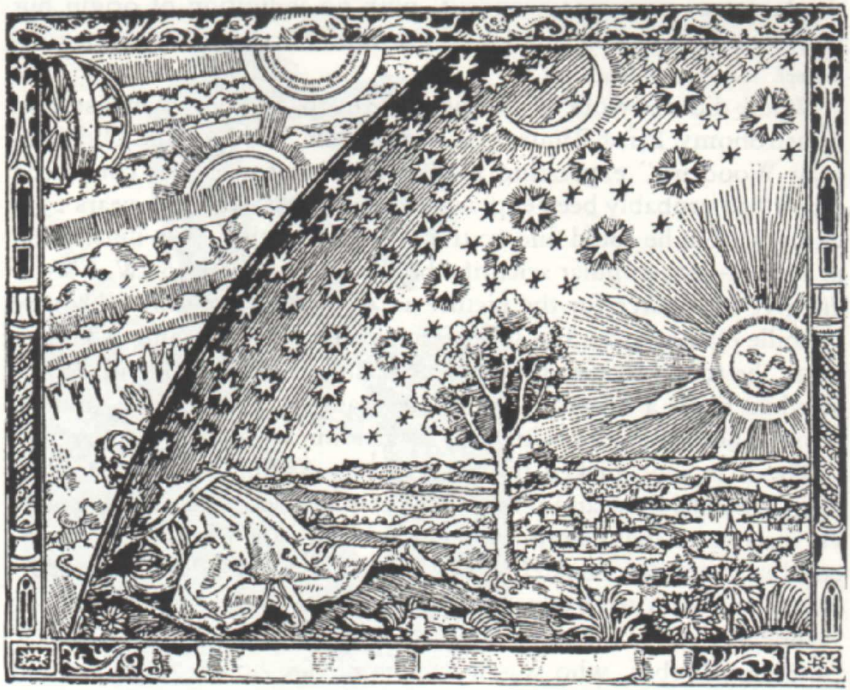}
\end{center}
\caption{\small The medieval view of a finite universe [woodcut by an unknown
  artist, first published by Camille Flammarion in
  {\em L'Atmosphere: M\'et\'eorologie Populaire\/} (Paris, 1888), p.~163]}
\label{fig1}
\end{figure}

The infinite extent of the standard model is at the very least a problematic
feature since it cannot be confirmed by experiment, even in principle.
One might be tempted to place it among the other theoretical infinities
that are presently tolerated in physics, such as those of quantum field theory.
In this work, we take a different approach and simply drop the assumption that
the material contents of the universe go on forever.  A large but finite
universe (say, millions of Hubble distances) should certainly be
observationally indistinguishable from one of infinite extent, at least for
an observer who is not too near the edge (figure~\ref{fig1}).
Indeed, it seems obvious that we can only hope to place a lower limit on the
universe's spatial extent.

\begin{figure}[t!]
\begin{center}
\includegraphics[width=75mm]{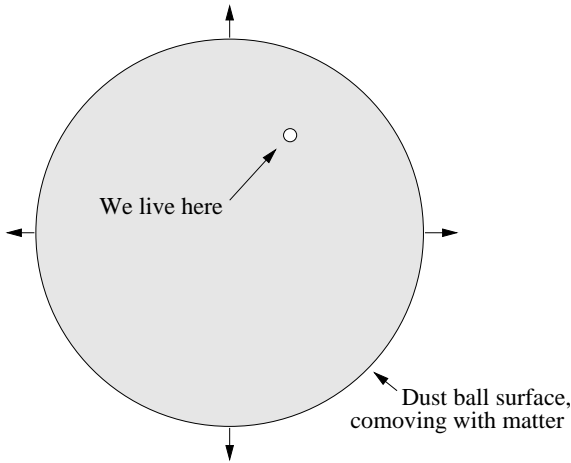}
\end{center}
\caption{\small The general features of our finite universe model.  The dust
   ball consists of cold matter and dark energy, and expands into an empty,
   time-independent exterior whose nature is to be determined.}
\label{fig2}
\end{figure}

The model of the universe that we develop below is consistent with the same
observational data that support the standard model in which the matter has an
infinite extent; it is a spherical dust ball surrounded by an empty exterior
space (figure~\ref{fig2}).  Both regions contain dark energy, represented
by a cosmological constant.  The dust ball is taken to have a standard FRW
geometry.  While this dust ball is of finite extent, the exterior space has no
boundary.  The geometry of the exterior is uniquely determined by that of the
dust ball, and is described by a metric first found by Kottler.  This is often
called a Schwarzschild-de~Sitter geometry because its metric approaches the
Schwarzschild metric for small radii and the static de~Sitter metric for large
radii.  The exterior has some remarkable features, prime among which is that
it harbors a naked singularity that fills all of 3-space before the big bang.
However, consideration of the earlier inflationary and radiation-dominated eras
leads us to believe that this singularity should probably be viewed as a
phenomenological representation for a small region of very heavy vacuum.

We note that this dust-ball universe is effectively a finite or truncated
version of the standard $\Lambda$CDM model, and should not be confused with
finite but topologically non-trivial models such as the Poincar\'e
dodecahedron.

The paper is organized as follows.  In section~2 we describe the geometry of
the dust ball and obtain lower limits on its total mass and size.  These limits
are determined by the fact that we do not at present see any obvious edge;
the mass limit is about $5 \times 10^{23}\MSUN$ and the size limit is about
50~Gly.  Section~3 contains comments on the possibility of testing the model; 
specifically, if we are close enough to the edge, there may be a ``cold spot''
detectable in the cosmic microwave background (CMB).  In section~4 we obtain the
metric for the exterior, which is completely determined by that of the dust
ball and the demands of spherical symmetry and time independence.  This is
most conveniently done in \PG\ (PG) coordinates, but we also give the exterior
metric in FRW-like and Schwarzschild coordinates.  In section~5 we transform the
metric for the exterior to conformal form, discussing the nature of the
exterior geometry and the picture of cosmogenesis that it suggests, wherein
the material contents of our universe explode from a 3-space-filling
singularity.  The emphasis of this work is on the present cosmic era, but in
section~6 we briefly sketch an approach to the early universe, incorporating
simple inflationary and radiation-dominated eras into the model; one amusing
result is that the singularity noted above is replaced by a region of heavy
vacuum.  Much further work could be done on the early finite universe.

We emphasize that the discussion of the dust ball and the mass and size limits
in sections~2 and 3 are well-founded extensions of the general-relativistic
standard cosmological model, and are self-contained.  The discussions of the
exterior, the early universe and cosmogenesis in later sections are more
speculative.  Altogether, we use six different coordinate systems in
obtaining our results; this is an amusingly large number but probably
not a record in the field.

\section{The universe in the dust ball}

We will assume that the standard cosmological model (defined here as flat
$\Lambda$CDM) is correct in its essential features.  Thus we take the interior
of the dust ball in figure~\ref{fig2} to have a spatially flat FRW metric,
and to be dominated by dark energy, represented by the cosmological constant,
and dust-like matter.  However, we truncate the matter at a comoving radius
$\udb$ to form a finite sphere, as shown in figure~\ref{fig2}.  The main
result of this section will be a lower limit on the mass and size of the
dust ball.  We work in FRW coordinates with both cosmic and conformal time
coordinates.

The metric of the dust ball in FRW coordinates is given in standard form (with
$c=1$) by
\be
ds^2 = dt^2 - a^2(t) [\,du^2 + u^2 (d\theta^2 + \sin^2\!\theta\,d\phi^2\,)]
   \; , \;\;\; u < \udb \; .
\label{ron2.1}
\ee
We use a dimensionless radial coordinate $u$ so that the scale function
$a(t)$ has the dimension of a distance; $\udb$ is the comoving radius of the
dust ball.  The scale function must obey the Friedmann equation, which results
from the Einstein field equations for the FRW metric, and reads
\be
\dot{a}^2 = \frac{{\cal C}}{a} + \frac{a^2}{\Rd^{\,2}} \; .
\label{friedmann}
\ee
Here ${\cal C}$ is a dimensionless constant of integration, and $\Rd$ is the
de~Sitter radius, which is related to the cosmological constant by 
$\Rd=\sqrt{3/\Lambda}$ and equal to about 16~Gly in the standard $\Lambda$CDM
model.  The solution appropriate to the present era is \cite{AO05}
\be
a(t) = ({\cal C}\Rd^{\,2})^{1/3} \sinh^{2/3} (3t/2\Rd) \; ,
\label{ron2.3}
\ee
and the present time is about 14~Gyr.

\begin{figure}[t!]
\begin{center}
\includegraphics[width=100mm]{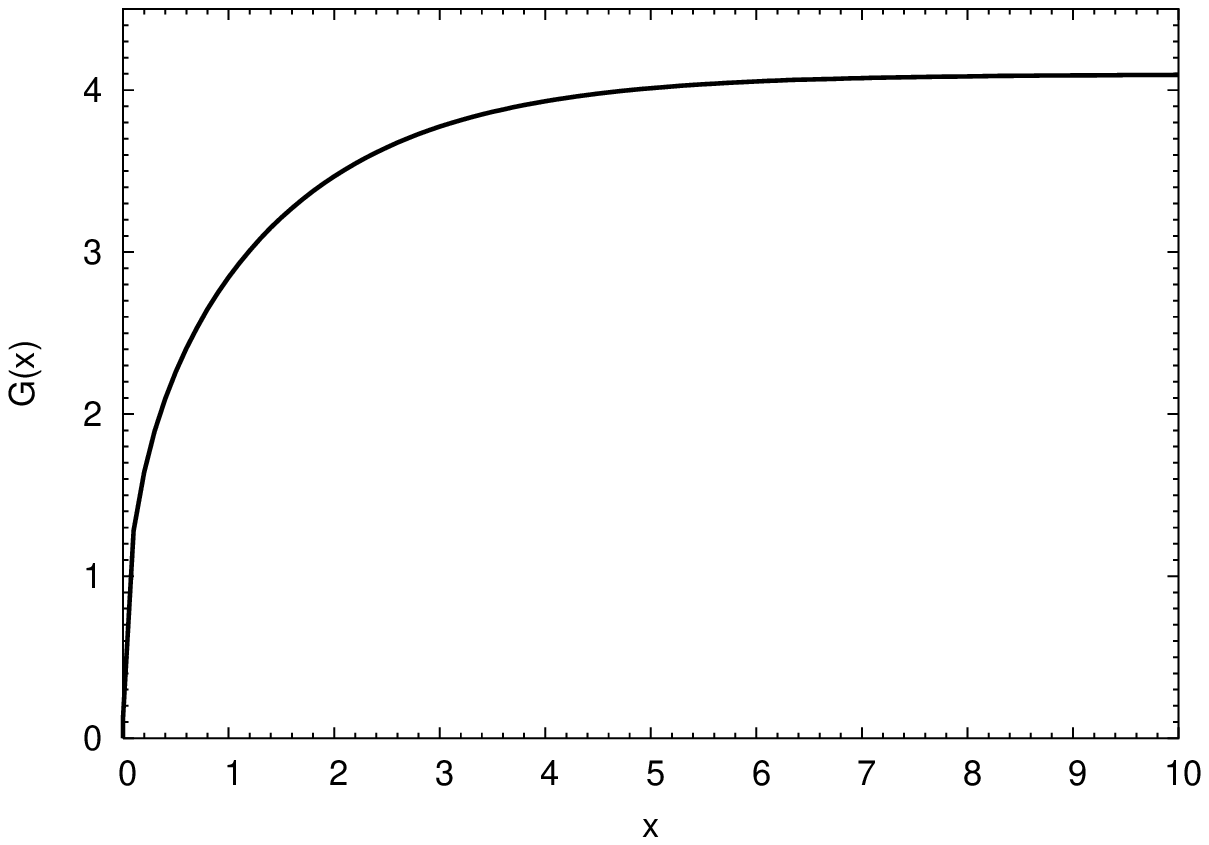}
\end{center}
\caption{\small Plot of the function
   ${\cal G}(x)=\int_0^x\sinh^{-2/3}(x^{\prime})\,dx^{\prime}$.}
\label{fig3}
\end{figure}

\begin{figure}[t!]
\begin{center}
\includegraphics[width=105mm]{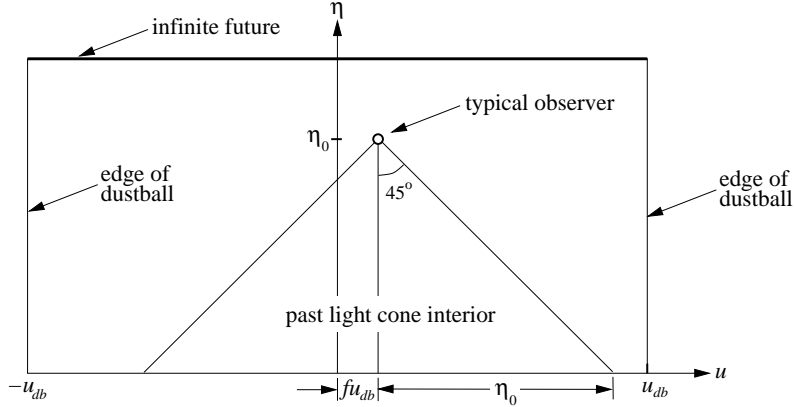}
\end{center}
\caption{\small A cut through the conformal FRW coordinate system, showing
   the observer's position and past light cone inside the dust ball.}
\label{fig4}
\end{figure}

To obtain a lower limit on the mass and size of the dust ball, we note that
no edge has been seen.  Thus our past light cone must lie entirely inside the
dust ball.  Analysis of the past light cone is most clearly done using
conformal time, wherein the metric is proportional to the Lorentz metric and
the light cone is the same 45$^{\circ}$ region as in special relativity.
Conformal time is defined by
\be
d\eta = dt/a(t) \; ,
\label{conformaltime}
\ee
in terms of which the metric becomes
\be
ds^2 = a^2(t) [\,d\eta^2 - du^2 - u^2 (d\theta^2 + \sin^2\!\theta\,d\phi^2\,)]
   \; , \;\;\; u < \udb \; .
\ee
The solution of eq.~(\ref{conformaltime}) is
\be
\eta = \frac{2}{3} \left( \frac{\Rd}{{\cal C}} \right)^{\!\!\!1/3} \!\!
   {\cal G}\!\left(\frac{3t}{2\Rd}\right) \; , \; \; \;
   {\cal G}\!\left(\frac{3t}{2\Rd}\right) \equiv \int_0^{\,3t/2\Rd} \!\!
   \frac{dx}{\sinh^{2/3}(x)} \; .
\label{Gdefn}
\ee
We have here chosen the constant of integration so that $t=0$ corresponds
to $\eta=0$.  The function ${\cal G}$ may be obtained numerically and is
plotted in figure~\ref{fig3}.
In particular, for the present time its value is about 3.2, so
\be
\eta_{\,0} = \frac{2}{3} \left( \frac{\Rd}{{\cal C}} \right)^{\!\!\!1/3}
   \!\! 3.2 \; , \;\;\; t_0 = \mbox{14 Gyr} \; .
\label{ron2.7}
\ee
For the infinite future, ${\cal G}$ may be calculated exactly in terms of
gamma functions, with the result
\be
{\cal G}(\infty) = \frac{\Gamma(1/6) \, \Gamma(1/3)}{2\Gamma(1/2)} = 4.2 \; ,
   \;\;\; \etainf = \frac{2}{3} \left( \frac{\Rd}{{\cal C}} \right)^{\!\!1/3}
   \!\!\!\! 4.2 \; , \;\;\; t = \infty \; .
\ee
We now require that our past light cone lie entirely inside the dust ball, as
shown in figure~\ref{fig4}.  It follows that
\be
\eta_{\,0} + f\udb \leqslant \udb \; ,
\label{ron2.9}
\ee
where $f$ is our fractional displacement from the center of the dust ball.
Eqs.~(\ref{ron2.7}) and (\ref{ron2.9}) together imply a limit on the radius of
\be
\frac{\eta_{\,0}}{(1-f)} = \frac{2}{3} \left( \frac{\Rd}{{\cal C}}
   \right)^{\!\!1/3} \frac{3.2}{(1-f)} \leqslant \udb \; .
\label{ron2.10}
\ee
Eq.~(\ref{ron2.10}) involves the physically unobservable quantities
${\cal C}$ and $\udb$; we prefer to express the inequality in terms of
physical quantities.  To do so, we first relate the integration constant
${\cal C}$ to the matter density using the $0,0$ component of the field
equations, which is
\be
8\pi G \rho = -\Lambda + 3 (\dot{a}^2/a^2) \; .
\ee
Comparison of this with the Friedmann equation~(\ref{friedmann}) gives
${\cal C}$ as
\be
{\cal C} = (8\pi G/3) \rho a^3 \; .
\label{ron2.12}
\ee
Eq.~(\ref{ron2.12}) allows us to calculate the total dust mass and
express it in terms of ${\cal C}$.  Specifically, we calculate twice the
geometric mass of the dust ball as
\be
2\mun = 2G\MUN = (8\pi G/3)\rho\,(a\udb)^3 = {\cal C}\udb^3 \; , \;\;\;
   {\cal C} = 2\mun / \udb^3 \; .
\label{ron2.13}
\ee
We will verify later that $2\mun$ plays the role of the Schwarzschild
radius of the dust ball, as viewed from the exterior.  Substituting
eq.~(\ref{ron2.13}) into the constraint~(\ref{ron2.10}), we obtain the
following limit on the dust ball's Schwarzschild radius in terms of the
de~Sitter radius:
\be
\frac{2}{3} \left( \frac{\Rd}{2\mun} \right)^{\!\!\!1/3} \!\!
   \frac{3.2}{(1-f)} \leqslant 1 \; , \;\;\;
   2\mun \geqslant \left( \frac{2}{3} \, \frac{3.2}{(1-f)} \right)^{\!\!3}
   \! \Rd \; .
\label{ron2.14}
\ee

One reasonable way to select a value for the parameter $f$ is to assume
that we are at a median position inside the dust ball; specifically, that
half the volume of the dust ball lies inside our radial position and half
outside.  This ``Copernican'' assumption implies that $f=(1/2)^{1/3}$, for
which eq.~(\ref{ron2.14}) gives
\be
2\mun \geqslant 1100 \Rd = 18,000 \mbox{ Gly} \;\;\;\;\;\;
  \mbox{ (median position)} \; .
\label{ron2.15}
\ee
Alternatively, we may be more conservative and assume only that our position
lies outside the central 5\% of the dust-ball volume.  That is, with 95\%
confidence we may say that $f\geqslant(.05)^{1/3}$, which gives
\be
2\mun \geqslant 39 \Rd = 620 \mbox{ Gly} \;\;\;\;\;\;
   \mbox{ (95\% confidence level)} \; .
\label{ron2.16}
\ee
The most conservative bound possible, of course, occurs for $f=0$, in which
case
\be
2\mun \geqslant 9.7 \Rd = 160 \mbox{ Gly} \;\;\;\;\;\;
   \mbox{ (lowest bound)} \; .
\label{ron2.17}
\ee
Since the Sun has a Schwarzschild radius of 2.9~km$=1.0\times10^{-5}$~ls,
eq.~(\ref{ron2.17}) translates into a lower bound on the mass of the
universe in solar units of
\be
\MUN \geqslant 5.1 \times 10^{23} \MSUN \; ,
\label{ron2.18}
\ee
where $\Rd=16$~Gly as before.

A lower bound on the present physical radius of the dust ball may be obtained
from eqs.~(\ref{ron2.3}) and (\ref{ron2.10}) with $f=0$ as 
\bea
\udb\,a(t) & \geqslant & \frac{2}{3} \left( \frac{\Rd}{{\cal C}}
   \right)^{\!\!1/3} \!\! \frac{3.2}{(1-f)} \, ({\cal C}\Rd^{\,2})^{1/3}
   \sinh^{2/3} (3t/2\Rd) \nonumber \\
& = & \left[ \frac{2}{3} \, \frac{3.2}{(1-f)} \sinh^{2/3} (3t/2\Rd) \right]
   \! \Rd \geqslant 3.1 \Rd = 49 \mbox{ Gly} \; .
\label{ron2.19}
\eea
As expected, this lower limit is considerably larger than the present Hubble
distance of about 14~Gly.  Eqs.~(\ref{ron2.18}) and (\ref{ron2.19}) are very
conservative; higher bounds would obviously result from using the median
assumption~(\ref{ron2.15}) or the 95\% confidence-level
assumption~(\ref{ron2.16}).

\section{Observational implications}

The scenario described above becomes experimentally testable if the size
of the dust ball ($a\udb$) is not too large or if we are sufficiently near
its edge.  In this case, the edge of our observable universe, which we
identify with the surface of last scattering, may be near the edge of
the dust ball.  It is likely that the CMB temperature would decrease to zero
(or a very low value) in some unknown way across a relatively thin shell
(of characteristic width $\varepsilon$) around the dust ball, which would
produce a cold spot in the CMB.  The situation is depicted in
figure~{\ref{fig5}, which might be compared with figure~\ref{fig2}.

\begin{figure}[t!]
\begin{center}
\includegraphics[width=90mm]{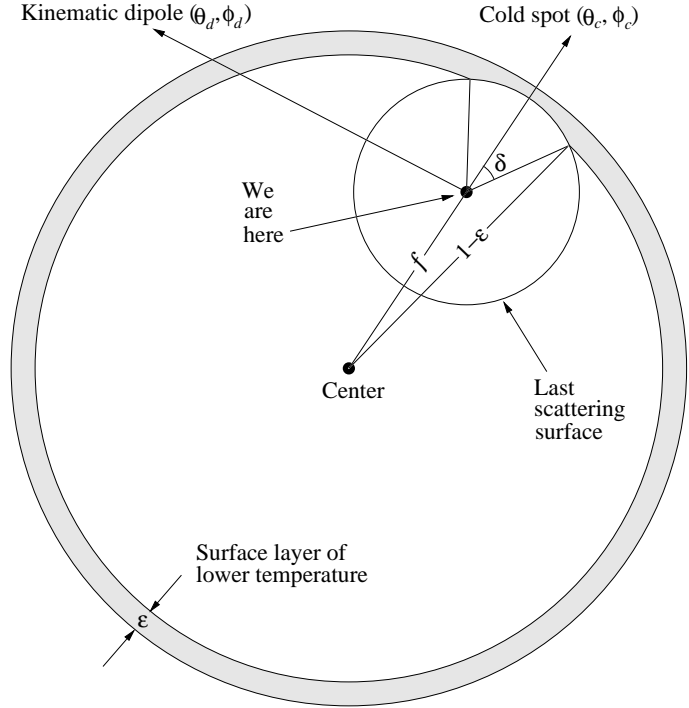}
\end{center}
\caption{\small Geometry of the finite dust-ball universe.
Distances are expressed as fractions of the dust-ball radius $\udb$.
The last scattering surface may protrude into the surface layer, thereby
producing a cold spot in the CMB.}
\label{fig5}
\end{figure}

After the removal of the galactic and other non-cosmological foregrounds,
the CMB temperature may be written in the form
\be
T = T_0 + T_1 + \Tm + \Tcs \; ,
\label{CMBsky}
\ee
where
\bea
T_0 & = & 2.73 \mbox{ K} \;\;\;\;\;\; \mbox{(monopole)} \nonumber \\
T_1 & = & T_0 \, v_d \cos\thetad = {\cal O}(10^{-3}\mbox{K}) \;\;\;\;\;\;
   \mbox{(kinematic dipole)} \nonumber \\
\Tm & = & {\cal O}(10^{-5}\mbox{K}) \;\;\;\;\;\;
   \mbox{(higher-order multipoles)} \nonumber \\
\Tcs & = & \mbox{ ?} \;\;\;\;\;\; \mbox{(cold spot)} \; . \nonumber
\eea
Here the Sun moves at speed $v_d$ in the direction $(\thetad,\phi_d)$ with 
respect to the CMB.  The term $\Tm$ contains the usual Harrison-Zeldovich
spectrum of higher-order temperature fluctuations detected by, for example,
the WMAP satellite, and $\Tcs$ is the cold spot we wish to discuss.

Assuming that the cold spot is not aligned with the Doppler dipole, the fact
that it has not already made itself obvious in CMB maps implies that its
magnitude is less than that of the dipole, $|\Tcs|\lesssim 10^{-3}$K.
Moreover, if $|\Tcs|\ll10^{-5}$K, then it is probably not possible to see
the effect on the CMB.  The interesting situation is therefore that in which
$|\Tcs|$ lies between approximately $10^{-6}$~K and $10^{-3}$~K.

The cold spot temperature, a function $\Tcs(\phi,\theta)$ of angular position,
will also depend on the characteristic angular width $\delta$, our fractional
distance $f$ from the center of the dust ball, and the relative size
$\varepsilon$ of the region of temperature decrease.  The angular width has
an upper limit that occurs when our line of sight just reaches the empty
exterior; it is given by
\be
f^2 + (1-f)^2 - 2f(1-f)\cos(\pi-\delta) = (1-\varepsilon)^2 \;\;\;\;\;\;
   \mbox{(upper limit on $\delta$)} \; ,
\ee
or, assuming that $\varepsilon\ll 1$,
\be
\delta^2 \leqslant 2\varepsilon/f(1-f) \;\;\;\;\;\;
   \mbox{(approximate upper limit on $\delta$)} \; .
\label{def}
\ee 
A lower limit occurs when our line of sight just reaches the shell of 
decreasing temperature, and is of course $\delta=0$.

A simple and reasonable choice for a parametric cold spot function is the
Gaussian distribution:
\begin{equation}
\Tcs=-\Tcso \exp(-\Delta\theta^2/2\delta^2) \; ,
\end{equation}
where $\Delta\theta$ is angular distance from the center of the cold spot.
As discussed above, we take the magnitude $\Tcso$ of the cold spot to lie
between $10^{-6}$K and $10^{-3}$K in cases of interest.  The most direct
approach to testing the finite-universe idea would be to fit eq.~(\ref{CMBsky})
to the CMB temperature data after cleaning the latter of foreground
contamination --- but not correcting for the kinematic dipole, which would
in general interfere with the desired cold spot signal.  In principle the fit
(not including the higher-order fluctuations) would involve seven parameters:
the magnitude and angular direction of the Doppler dipole
($v_d,\thetad,\phid$) and the magnitude, width and angular direction of the
cold spot ($\Tcso,\delta,\thetac,\phic$).

More sophisticated search techniques have recently led to several claims of
non-Gaussianity in the WMAP data \cite{Chi03,Lar05,Toj05,Ber05}, parts of
which are likely local in origin since they exhibit correlations with the
ecliptic plane \cite{Eri03,Eri04,Han04a,Han04b,Han04c,Cop05,Fre05} and the
diffuse $\gamma$-ray background as measured by the EGRET satellite
\cite{Liu05}.  Of special interest for our purposes is a growing body of
statistical analysis suggesting that the WMAP 1-year data is {\em nearly\/}
Gaussian --- but for a single cold spot in the direction
$(l,b)=(209^{\circ},-57^{\circ})$ that is well-fit by a Gaussian function of
width $\delta=4^{\circ}$ and magnitude $\Tcso=73~\mu$K
\cite{Vie03,Cru04,McE05a,Cay05}.  This spot appears to be incompatible with
either a thermal Sunyaev-Zeldovich effect or incomplete galactic foreground
subtraction \cite{Cru06}, leading theorists to speculate on such possible
causes as large-scale inhomogeneity \cite{Tom05}, Bianchi-type models with
nonzero shear or rotation \cite{McE05b} and homogeneous, spherically-symmetric
local voids \cite{Ino06}.  Alternatively, we note that the
reported value of $\Tcso$ lies within the range of interest for a universe
that is standard in every way except that it is of finite, rather than
infinite spatial extent .  Accepting the reported value of $\delta$
and adopting the median or most likely value of $f=(1/2)^{1/3}$, we find
from eq.~(\ref{def}) that the fractional edge thickness $\varepsilon$ in
such a model is less than about $4.0\times10^{-4}$.

Some cosmologists believe there is independent evidence for a closed and finite
universe in the suppression (relative to the dipole) of the lowest-order
multipoles in the power spectrum of CMB fluctuations \cite{Efs03,Uza03}.
In particular, the quadrupole moment is about seven times weaker than expected
on the basis of a flat $\Lambda$CDM model, and the octupole is only about
72\% of its expected value \cite{Lum05}.  While this suppression may be
an artifact of cosmic variance, it can also arise if the universe simply does
not have enough room for the longest-wavelength fluctuations.  Non-trivial
topology is not required for such suppression; it is only necessary that the
initial fluctuation spectrum truncate at around the curvature scale of the
universe \cite{Efs03}.  Such an effect might allow one to perform an
independent experimental test of the finite-universe model by re-computing
the low-order CMB multipole moments $C_{\ell}$, using eq.~(\ref{CMBsky})
and standard techniques.
%...normalizing them to the high-$\ell$ end of the power spectrum for
% reference.  The $C_{\ell}$ are found from
%\begin{equation}
%C_{\ell} = 2\pi \int_0^{\pi} {\cal C}(\theta) P_{\ell}(\cos\theta) \sin\theta
%   d\theta \; ,
%\end{equation}
%where the $P_{\ell}(x)$ are standard Legendre polynomials and
%${\cal C}(\theta)$ is the correlation coefficient for the CMB temperature
%$T(\vec{\Omega})$ as given by eq.~(\ref{CMBsky}):
%\begin{equation}
%{\cal C}(\theta) = \frac{\mbox{Cov}[T(\vec{\Omega}_1),T(\vec{\Omega}_2)]}
%   {\mbox{Var}[T(\vec{\Omega}_1)]} \; .
%\end{equation}
%Here $\theta$ is the angle between two directions $\vec{\Omega}_1$ and
%$\vec{\Omega}_2$ in the sky.
It would be of considerable interest to see if our finite-universe model
produced the observed suppression of low-order multipoles; this would involve
the thermodynamcs of the early universe, which we will briefly discuss in
section~6.

\section{Outside the dust ball}

The dust ball in our model universe has a spatially flat FRW metric, and
according to section~2 its Schwarzschild radius is much larger than its
de~Sitter radius, $2\mun/\Rd\gg1$.  To study the exterior, we naturally
assume that its metric is spherically symmetric and time-independent.
Of course, it must also match the dust ball on its surface.  These
assumptions uniquely determine the exterior geometry.

The spherical symmetry and time independence of the exterior lead to a
standard Schwarzschild form for the metric, which we write in the form
\be
ds^2 = (1-v^2)\,dt_s^2 - \frac{dr^2}{(1-w^2)} - r^2 d\Omega^{\,2} \; ,
   \;\;\;\;\;\; v=v(r), w=w(r) \; .
\label{ron4.1}
\ee
The mathematical problem is to match this to the metric of the dust ball at
the surface, and thereby obtain the functions $v(r)$ and $w(r)$.  This is
particularly easy in \PG\ (PG) coordinates, which use both the FRW cosmic time
and the Schwarzschild radial marker, thereby interpolating effectively between
the two sets of coordinates.  We therefore transform both the dust-ball
metric~(\ref{ron2.1}), with eq.~(\ref{ron2.3}), and the exterior 
metric~(\ref{ron4.1}) to PG coordinates.  For the exterior
we introduce the PG time $t$ as
\be
t_s = t - {\cal P}(r) \; ,
\ee
and obtain the metric in PG form \cite{Pai21,Gul22}:
\be
ds^2 = (1-v^2)\,dt^2 \pm 2 v \, dr \, dt -dr^2 - r^2 d\Omega^{\,2} \; ,
\label{ron4.3}
\ee
provided that we choose $w(r)=v(r)$ and require that ${\cal P}$ satisfy
\be
{\cal P}^{\prime} = \pm \frac{v}{1-v^2} \; .
\ee
For our purposes, it is not necessary to solve explicitly for ${\cal P}$.
For the dust ball, we transform the dimensionless FRW radial coordinate to a
new radial coordinate by
\be
r = a(t) u 
\label{ron4.5}
\ee
and obtain
\be
ds^2 = [1-(r\dot{a}/a)^2]\,dt^2 + 2\,(r\dot{a}/a) \, drdt - dr^2 - r^2
   d\Omega^{\,2} \; ,
\label{ron4.6}
\ee
where $a(t)$ is given explicitly for the present era in eq.~(\ref{ron2.3}),
and the overdot indicates differentiation with respect to time $t$.  Both the
exterior metric~(\ref{ron4.3}) and dust-ball metric~(\ref{ron4.6}) now have
the same form, which makes the PG coordinates very convenient.  It only
remains to match the functions $v$ and $r\dot{a}/a$ at the dust-ball surface.

To determine the function $v(r)$ of the exterior, we first demand that the
motion of the dust-ball surface, which we denote by $r_s(t)$, be consistent
with the comoving interior material at the surface, or
\be
r_s(t)=\udb a(t) \;\;\;\;\;\; \mbox{(surface)} \; .
\label{ron4.7}
\ee
We also demand that the dust-ball and exterior metrics match at the surface, or
\be
v = r_s(t) \dot{a}/a = \udb \dot{a} \;\;\;\;\;\; \mbox{(surface)} \; .
\label{ron4.8}
\ee
We substitute for $\dot{a}$ from eq.~(\ref{ron2.3}) to obtain
eq.~(\ref{ron4.8}) in the form:
\be
v = \udb \left( \frac{{\cal C}}{\Rd} \right)^{\!\!\!1/3} \!\!\!
   \frac{\cosh(3t/2\Rd)}{\sinh^{1/3}(3t/2\Rd)} \;\;\;\;\;\;
   \mbox{(surface)} \; .
\label{ron4.9}
\ee
Next, using eqs.~(\ref{ron2.3}) and (\ref{ron4.7}), we relate $t$ to $r_s$
on the surface:
\be
\sinh (3t/2\Rd) = \frac{r_s^{3/2}}{\udb^{3/2}(\Rd^{\,2}\,{\cal C})^{1/2}}
   \;\;\;\;\;\; \mbox{(surface)} \; .
\label{ron4.10}
\ee
Finally, we substitute from eq.~(\ref{ron4.10}) into eq.~(\ref{ron4.9})
to obtain
\be
v = \sqrt{ \frac{{\cal C}\,\udb^3}{r} + \frac{r^2}{\Rd^{\,2}} } =
   \sqrt{ \frac{2\mun}{r} + \frac{r^2}{\Rd^{\,2}} } \; .
\label{ron4.11}
\ee
Since $v$ is a function of only $r$, Eq.~(\ref{ron4.11}) must hold throughout
the exterior and not just on the surface of the dust ball.  We have therefore
dropped the subscript on $r$; this gives the metric of the exterior.  
Explicitly, in PG coordinates,
\be
ds^2 = \left( 1 - \frac{2\mun}{r} - \frac{r^2}{\Rd^{\,2}} \right)\,dt^2 +
   2 \sqrt{ \frac{2\mun}{r} + \frac{r^2}{\Rd^{\,2}} } \, dr \, dt -
   dr^2 - r^2 d\Omega^{\,2} \; ,
\label{ron4.12}
\ee
where the plus sign of the cross term is appropriate to an expanding dust
ball.

It may be verified that the geodesic equation of motion of a zero-energy test
particle moving outward in the exterior geometry~(\ref{ron4.3}) is given by
\be
dr/dt = v(r) \; .
\label{ron4.13}
\ee
This is consistent with eqs.~(\ref{ron4.7}) and (\ref{ron4.8}) for the
dust-ball surface, and indeed guarantees that the interior and exterior 
geometries are consistent.

In Schwarzschild coordinates, the exterior metric has the form
\be
ds^2 = \left( 1 - \frac{2\mun}{r} - \frac{r^2}{\Rd^{\,2}} \right)\,dt_s^2 -
   \left( 1 - \frac{2\mun}{r} - \frac{r^2}{\Rd^{\,2}} \right)^{\!\!-1}
   \!\!\!\! dr^2 - r^2 d\Omega^{\,2} \; .
\ee
This metric was first obtained as a solution of the vacuum field equations
by Kottler in 1918 \cite{Kot18} and Weyl in 1919 \cite{Wey19}, but is now
often referred to as the Schwarzschild-de~Sitter metric since it obviously
combines the Schwarzschild and static de~Sitter metrics.

We emphasize that in obtaining the exterior metric, we used only the 
assumptions of spherical symmetry and time independence, and the metric
inside the dust ball; we did not impose the field equations.  The metric is a
vacuum solution with a cosmological constant, but we did not force it to be so.

\begin{figure}[t!]
\begin{center}
\includegraphics[width=55mm]{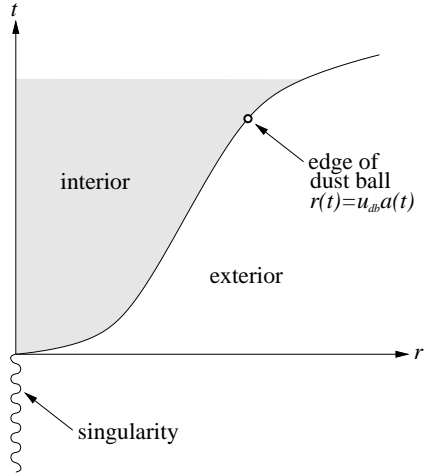}
\end{center}
\caption{\small Evolution of the dust-ball and exterior regions in time, in
   \PG\ coordinates.}
\label{fig6}
\end{figure}

In figure~\ref{fig6} we show the dust-ball and exterior regions of our
model universe in PG coordinates.  The equation of the surface is given by
eq.~(\ref{ron4.7}), with $a(t)$ given in eq.~(\ref{ron2.3}).  Note that prior
to $t=0$, there is a singularity at $r=0$, as is evident from
eq.~(\ref{ron4.12}).

The exterior metric may also be expressed in FRW-like coordinates with the
use of the transformation~(\ref{ron4.5}); this results in
\be
ds^2 = dt^2 - {\cal D}^2 - a^2(t) u^2 d\Omega^{\,2} \; ,
\ee
where
\bea
{\cal D} = & & \hspace{-6mm} \left( \frac{{\cal C}}{\Rd\sinh(3t/2\Rd)}
   \right)^{\!\!1/3} \!\! \left\{ u \left[ \sqrt{ \cosh^2(3t/2\Rd) + 
   (\udb^3/u^3-1) } - \cosh(3t/2\Rd) \right]\,dt \right. \nonumber \\
   & & \left. \hspace{-7mm} \frac{ }{ } - \Rd \sinh(3t/2\Rd) du \right\}
   \; .
\eea
Clearly, this joins smoothly to the dust-ball FRW metric at the surface
$u=\udb$, as is evident from eqs.~(\ref{ron2.1}) and (\ref{ron2.3}).

As a bonus, we point out that the above results apply also to the problem of
the gravitational collapse of a uniform dust ball in a de~Sitter background
space; it is only necessary to reverse the sign of the PG time coordinate, or
equivalently the sign of the cross term in the PG form of the metric 
\cite{Adl05}.

In the next section, we will study the Kottler geometry in more detail, showing
that it has peculiar and interesting properties when $2\mun\gg\Rd$, which is
the relevant case for our model.

\section{Geometric nature of the exterior}

In this section, we will use the Kottler metric as expressed in
eq.~(\ref{ron4.12}) to investigate the geometrical nature of the exterior
universe.  Specifically, we will show that both the Schwarzschild and 
PG ``time'' coordinates, while algebraically convenient, are not
acceptable time markers in the exterior space.  Also, we will see that
there is a rather interesting naked singularity at the origin for $t<0$.

Consider an observer at rest in the PG coordinates, with
$dr=d\theta=d\phi=0$, so that the line element is
\be
ds^2 = \left( 1 - \frac{2\mun}{r} - \frac{r^2}{\Rd^{\,2}} \right)\,dt^2 \; .
\label{ron5.1}
\ee
If the $g_{00}$ component of the metric (in brackets) is positive, then the
proper time interval $ds$ and the coordinate interval $dt$ are both real, so
$t$ is an acceptable time marker.  If $g_{00}$ is negative, then $t$ is not an
acceptable time marker.  The Kottler $g_{00}$ is plotted in
figure~\ref{fig7} for two cases.
\begin{figure}[t!]
\begin{center}
\includegraphics[width=75mm]{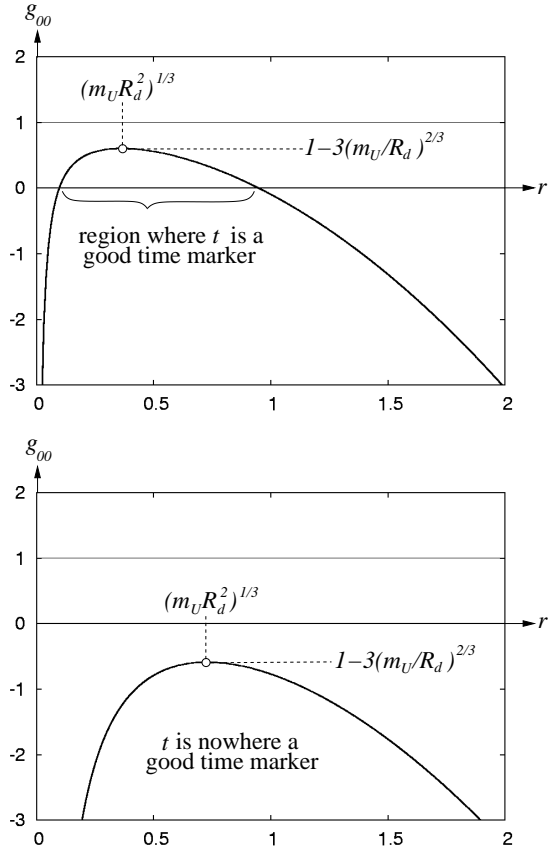}
\end{center}
\caption{\small Metric component $g_{00}$ and the Painlev\'e ``time''
   coordinate.
%  Plotted are the cases $\mun=\Rd/4\sqrt{27}$ (top) and
%   $\mun=2\Rd/\sqrt{27}$ (bottom) with $\Rd=1$.
   }
\label{fig7}
\end{figure}
In the case $2\mun < (2/\sqrt{27})\Rd$, it increases from minus infinity
at $r=0$, lies between zero and one over some range of $r$, and then decreases
to become negative again for large $r$ (figure~\ref{fig7}, top).  In the
range where $g_{00}$ is positive, the coordinate $t$ is a useful time marker
and we can interpret the metric as describing a Schwarzschild black hole
in a de~Sitter background.  Indeed, if the geometric mass is much less than the
de~Sitter radius, that is $2\mun\ll (2/\sqrt{27})\Rd$, then the coefficient
in eq.~(\ref{ron5.1}) is zero near $2\mun$ and $\Rd$, so the black-hole
radius is near $2\mun$.

However, in the case of our model, $2\mun\gg (2/\sqrt{27})\Rd$ and we see that
$g_{00}$ is always negative, so that $t$ is not a useful time marker anywhere in
space (figure~\ref{fig7}, bottom).  This situation is, of course, analogous
to that of the standard Schwarzschild black-hole interior, which is better
described using Kruskal-Szekeres coordinates than Schwarzschild coordinates.
Accordingly, we now transform to new radial and time coordinates in which the
metric is proportional, or conformal, to that of two-dimensional flat space
with Lorentz coordinates.  (In the rest of this section, we will deal only
with the $t,r$ subspace, with angular coordinates suppressed.) We will refer
to these as {\em conformal coordinates\/} for brevity.  Since the light cones
are the same as in two-dimensional special relativity, the causal structure
of the space-time is quite transparent.

The metric (with angular parts suppressed) is
\be
ds^2 = (1-v^2)\,dt^2 + 2 v \, dr \, dt - dr^2 \; ,
\ee
so that radial null lines, or light rays, are described by
\be
dt = \frac{dr}{(v \pm 1)} \;\;\;\;\;\; \mbox{(+ and -- null lines)} \; .
\label{ron5.3}
\ee
These null lines and some light cones are illustrated in figure~\ref{fig8}.
\begin{figure}[t!]
\begin{center}
\includegraphics[width=80mm]{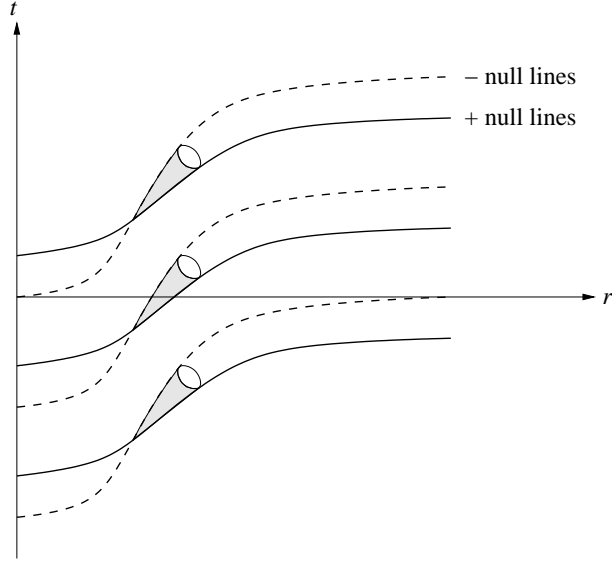}
\end{center}
\caption{\small Null lines and schematic light cones in the Painlev\'e 
   coordinates for a pure Kottler geometry.}
\label{fig8}
\end{figure}
Based on eq.~(\ref{ron5.3}), we introduce null coordinates defined by
\bea
& & \mu = t - \int_0^r \!\! \frac{dr^{\prime}}{v+1} \;\;\; , \;\;\;
   \lambda = \int_0^r \!\! \frac{dr^{\prime}}{v-1} - t \;\;\; , \nonumber \\
& & d\mu = dt - \frac{dr}{v+1} \;\;\; , \;\;\;
   d\lambda = \frac{dr}{v-1} - dt \; .
\label{ron5.4}
\eea
Note that $d\mu=0$ and $d\lambda=0$ correspond to the $+$ and $-$ null lines,
so that lines of constant $\mu$ or $\lambda$ represent light rays.  From
eqs.~(\ref{ron5.4}) we calculate the metric in the new coordinates to be
\be
ds^2 = (v^2 - 1)\,d\mu \, d\lambda \; ,
\ee
in which $v$ is to be considered an implicit function of $\mu$ and $\lambda$
from eqs.~(\ref{ron4.11}) and (\ref{ron5.4}).

To see the geometrical relation of the null coordinates to the PG coordinates,
we consider some special lines, with the results shown in figure~\ref{fig9}.
\begin{figure}[t!]
\begin{center}
\includegraphics[width=130mm]{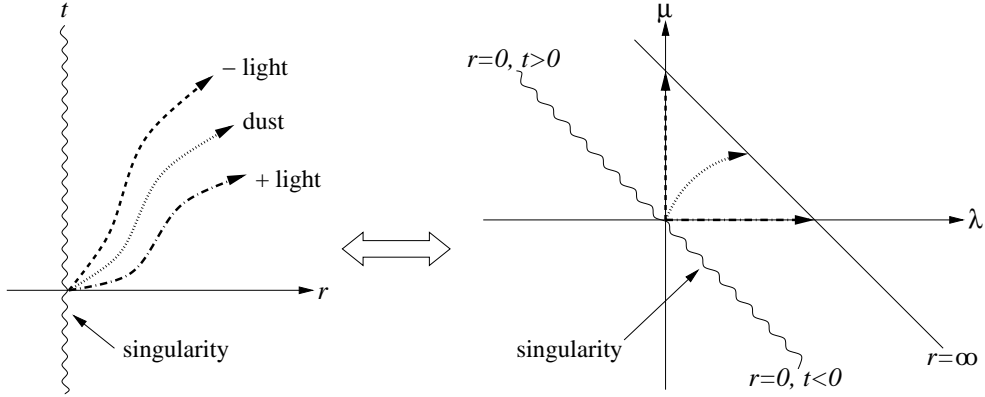}
\end{center}
\caption{\small Some corresponding trajectories and points in PG and null
   conformal coordinates.}
\label{fig9}
\end{figure}
The figure shows only the exterior Kottler geometry, with no dust ball.
Note that the combination
\be
\mu + \lambda = 2 \int_0^r \frac{dr^{\prime}}{v^2 - 1}
\ee
is independent of $t$.  In particular, the singular line $r=0$ corresponds to
$\mu+\lambda=0$, while the line $r=\infty$ corresponds to
\be
\mu + \lambda = 2 \int_0^{\infty} \!\!\! \frac{dr^{\prime}}{v^2 - 1} =
   2 \int_0^{\infty} \!\!\! \frac{dr^{\prime}}{ (2\mun/r^{\prime}) + 
   r^{\prime\,2}/\Rd^{\,2} - 1} \;\;\;\;\;\; (r=\infty) \; .
\ee
For $\Rd\ll 2\mun$, the function $v^2$ is always much larger than one, so that
the integral may be approximated by dropping the one in the denominator, with
the result that
\be
\mu + \lambda \approx 2 \Rd (\Rd/2\mun)^{1/3} \frac{2\pi}{3\sqrt{3}}
   \;\;\;\;\;\; (r=\infty) \; .
\label{ron5.7b}
\ee
From the inequality~(\ref{ron2.14}), eq.~(\ref{ron5.7b}) implies that
\be
\mu + \lambda \approx \left[ \frac{2\pi}{\sqrt{3}} \frac{(1-f)}{3.2}
   \right] \, \Rd \leqslant \left[ \frac{2\pi}{3.2\sqrt{3}} \right] \, \Rd =
   1.1\,\Rd \; .
\ee
That is, $\mu + \lambda$ is at most of order $\Rd$.  The entire physical space
lies between the diagonal lines shown in figure~\ref{fig9}.  The +~light ray
emitted from the origin corresponds to $\mu=0$ and positive $\lambda$, while
the --~light ray emitted from the origin corresponds to $\lambda=0$ and
positive $\mu$, as shown in the figure.  Finally, the trajectory of a
zero-energy dust particle emitted from the origin and obeying
eq.~(\ref{ron4.13}) corresponds to the parametric curve
\be
\mu = \int_0^r \frac{dr^{\prime}}{v} -
   \int_0^r \frac{dr^{\prime}}{v+1} =
   \int_0^r \frac{dr^{\prime}}{v^2+v} \;\;\; , \;\;\;
\lambda = \int_0^r \frac{dr^{\prime}}{v-1} -
   \int_0^r \frac{dr^{\prime}}{v} =
   \int_0^r \frac{dr^{\prime}}{v^2-v} \; ,
\ee
which is sketched in the figure.  Of course, this also describes the dust-ball
surface.

As our last change of coordinates, we rotate the $\mu,\lambda$ null
coordinates by $45^{\circ}$ to obtain Lorentz-like coordinates $\tau,\rho$
defined by
\bea
& & \tau = (\lambda + \mu)/2 = \int_0^r \!\!
   \frac{dr^{\prime}}{2\mun/r^{\prime} + r^{\prime\,2}/\Rd^{\,2} - 1}
   \; , \nonumber \\
& & \rho = (\lambda - \mu)/2 = \int_0^r \!\! 
   \frac{\sqrt{2\mun/r^{\prime} + r^{\prime\,2}/\Rd^{\,2}} \,
   dr^{\prime}}{2\mun/r^{\prime} + r^{\prime\,2}/\Rd^{\,2} - 1} - t \; .
\eea
In these coordinates, the metric is
\be
ds^2 = (v^2 - 1) (d\tau^2 - d\rho^2) \; .
\ee
The exterior space in these coordinates is shown in Fig~\ref{fig10}
with some special lines indicated, and with the dust-ball region excised.

\begin{figure}[t!]
\begin{center}
\includegraphics[width=65mm]{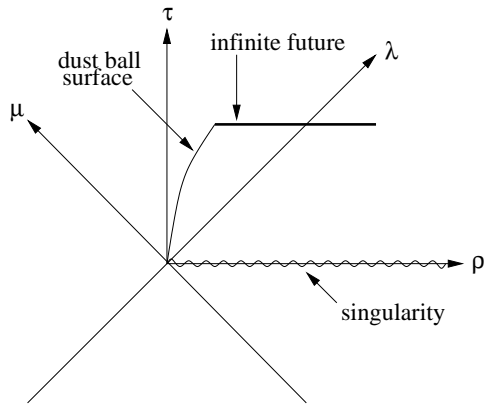}
\end{center}
\caption{\small The exterior region in Lorentz-like coordinates.  Only the
   region between the singularity and infinite future is physical, and
   similarly only the region from the dust-ball surface to infinite radial
   distance.}
\label{fig10}
\end{figure}

According to figure~\ref{fig10}, we can view our model universe as a dust
ball ejected from a singular hypersurface, which comprises all of 3-space
at the initial time $\tau=0$.  This singularity is analogous to the 
singularity of Schwarzschild geometry, which in Kruskal-Szekeres
coordinates forms a future spacelike barrier to light and particles inside
the black-hole surface.  Note that the surface of the dust ball can be
crossed by light and particles moving both inward and outward; that is,
the interior of the dust ball can communicate with the exterior, and
vice versa.

\section{The early universe}

The focus of this paper is on the present cosmic era and a universe 
dominated by dark energy and cold matter.  We have not worked out the dynamics
of the early universe in detail, but in this section we will briefly sketch
an overal cosmic history that includes radiation-dominated and inflationary
eras.  Before the time of matter-radiation equality, at about
$\tE\approx10^5$~yr, the universe was dominated by radiation and relativistic
or hot matter, with an approximate equation of state $p=\rho/3$.  The scale
function for hot matter is well known to take the form
\be
a(t) \propto \sinh^{1/2} (t/2\Rd) \;\;\;\;\;\; \mbox{(radiation era)} \; .
\label{ron6.1}
\ee
This may be compared to the scale function during the matter-dominated era,
eq.~(\ref{ron2.3}).  Both are singular at $t=0$.  During the
radiation-dominated era, $t\ll\Rd$ and the scale factor is well approximated
by a constant times $a\propto t^{1/2}$.  During the matter-dominated era,
the scale function~(\ref{ron2.3}) is well approximated by a constant times
$t^{2/3}$.

We can write the scale function for the early universe in a convenient and
well known approximate form, if we assume that the energy density is entirely
dominated by radiation (with $p=\rho/3$) before matter-radiation equality,
and matter (with $p=0$) after this time.  By matching the scale function and
its derivatives at the time of equality, $\tE$, we obtain the expressions
\bea
a(t) & \approx & ({\cal C}^{1/3}\,2^{2/3}\,\tE^{1/6})\,t^{1/2} \;\;\;\;\;\;
   \tI<t<\tE \;\;\;\;\;\; \mbox{(fireball)} \nonumber \\
a(t) & \approx & {\cal C}^{1/3}(3/2)^{2/3}(t+\tE/3)^{2/3} \;\;\;\;\;\;
   t>\tE \;\;\;\;\;\; \mbox{(early dust ball)} \; ,
\label{ron6.2}
\eea
where ${\cal C}$ is the same constant of integration used in section~2, and
$\tI$ is the time at which the inflationary era ends and the
radiation-dominated one begins.  We refer to the universe before $\tE$ as a
fireball, to distinguish it from the dust ball that follows.

The mathematical problem is then to match the early fireball geometry to the
exterior Kottler geometry, just as we matched the dust ball to the exterior
in section~4.  However this cannot be done in quite the same way, since there
is pressure in the fireball, and thus a pressure gradient at the surface.
The pressure gradient will eject material from the fireball surface, which
will cool by expansion and radiation to form a cooler shell surrounding the
ball.  Hence the surface region will have a lower temperature and pressure.
Indeed, for consistency with the time-independent exterior geometry obtained
for the present era in section~4, the outer layers of the fireball must be at
a reasonably low temperature and behave at least approximately like dust.

Clearly there is a wealth of further problems to be explored concerning the
radiation-dominated era, including the density and pressure and temperature
profiles of the fireball, as well as the effect of the finite size of the
fireball on the CMB discussed in section~3.  Such problems will probably 
require further and deeper use of the Einstein equations with the Einstein
tensor given below in eqs.~(\ref{Eequns}).

To summarize this part of the model: it seems plausible that the
radiation-dominated era does not significantly alter the qualitative cosmic
picture we have obtained for the present era; only the inner region of the
fireball needs to be changed before the epoch of matter-radiation equality,
as shown in the upper part of figure~\ref{fig11}.

\begin{figure}[t!]
\begin{center}
\includegraphics[width=60mm]{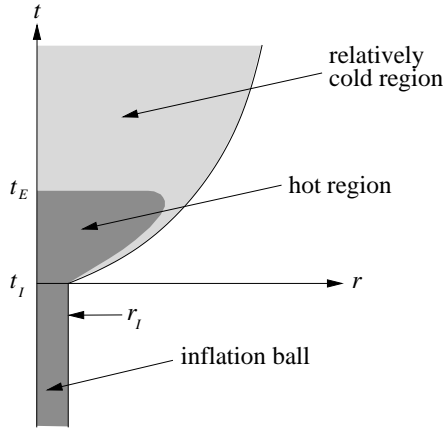}
\end{center}
\caption{\small Schematic sketch of the early universe, including a hot
   radiation-dominated region and an inflationary region.}
\label{fig11}
\end{figure}

We next consider an inflationary era preceding the radiation-dominated one,
as is currently popular.  Inflation eliminates the singularity at $t=0$ and
replaces it with a region of de~Sitter-like geometry.  The inflationary era
is usually modeled in terms of one or more scalar fields, but is often
described approximately with the use of a cosmological constant.  Our model
for this era is a ball of (static) de~Sitter space, which we call an
inflation ball, as shown in the lower part of figure~\ref{fig11}.

We first calculate the Schwarzschild radius of the fireball at the end of
inflation and beginning of the radiation-dominated era.  The density of the
matter and radiation in the fireball at this time is the difference between
the vacuum-energy density $\rI$ during inflation and that in the present
era ($\rO$).  We refer to all such matter as ponderable matter and write
its density as follows:
\be
\rpond = \rI - \rO = \frac{3}{8\pi G\RI^{\,2}} - \frac{3}{8\pi G\Rd^{\,2}} \; ,
\ee
where $\RI$ is the de~Sitter radius during inflation, a number typically taken
to be of order $10^{-26}$~m.  The Schwarzschild radius is obtained from this
density as
\be
2\mun = 2G\MUN = 2G \left( \frac{4\pi\ri^3}{3} \right) \rpond =
   \ri^3 \left( \frac{1}{\RI^{\,2}} - \frac{1}{\Rd^{\,2}} \right)\; ,
\label{ron6.4}
\ee
where $\ri$ denotes the initial radius of the fireball.

For the inflationary era, the scale function is an exponential, which we write
as
\be
a(t) = ({\cal C}^{1/3}\,2^{2/3}\,\tE^{1/6}\,\tI^{1/2})\,e^{(t-\tI)/\RI} \; ,
   \;\;\;\;\;\; t<\tI \; .
\ee
The coefficient here is chosen in such a way as to match the scale function
and its derivative at the beginning of the radiation-dominated era (time
$\tI$).  From the transformation~(\ref{ron4.5}), we find that the corresponding
PG metric function is
\be
v = (\dot{a}/a)r = r/\RI \; .
\label{ron6.6}
\ee
This represents static de~Sitter space.  We must now equate this function $v$
to the exterior function $v$ in eq.~(\ref{ron4.11}) at the surface of the
inflation ball.  Since both functions depend only on radius, the equality
must hold along a line of constant radius $\ri$ given by
\be
v = \sqrt{\frac{2\mun}{\ri} + \frac{\ri^2}{\Rd^{\,2}}} = \frac{\ri}{\RI} \; ,
   \;\;\;\;\;\; 2\mun = \ri^3 \left( \frac{1}{\RI^{\,2}} - \frac{1}{\Rd^{\,2}}
   \right) \; .
\label{ron6.7}
\ee
Remarkably, this is the same relation that we found in eq.~(\ref{ron6.4}),
which resulted from energy conservation.  The inflation ball thus has a 
constant radius, as shown in figure~\ref{fig11}.

The above matching of the inflation-ball geometry to the exterior is quite
unlike the matching that we used for the present era, in which the interior
and exterior metrics matched along a test-particle (or dust-particle)
geodesic.  The reason for the difference is that the inflation ball has
surface tension.  To see this, we note that the function $v$ is continuous at
the inflation-ball surface according to eq.~(\ref{ron6.7}), but that its
derivative is not.  The discontinuity in the derivative produces a singular
stress-energy tensor at the surface, which can be calculated from the
Einstein equations $T_{\mu}^{\nu}=-G_{\mu}^{\nu}/(8\pi G)$.  It is only
slightly tedious to calculate the Einstein tensor for the PG form of the
metric, with the results \cite{Adl05}
\bea
& & G_0^0 = -\frac{2vv^{\prime}}{r}-\frac{v^2}{r^2} \; , \;\;\;\;\;\;
   G_0^1 = \frac{2v\dot{v}}{r} \; , \nonumber \\
& & G_1^1 = -\frac{2vv^{\prime}}{r}-\frac{v^2}{r^2}-\frac{2\dot{v}}{r} \; ,
   \;\;\;\;\;\; G_2^2 = G_3^3 = -\frac{2vv^{\prime}}{r} - \frac{\dot{v}}{r} -
   \dot{v}^{\prime} - v^{\prime\prime}v - v^{\prime 2} \; .
\label{Eequns}
\eea
Here primes denote radial derivatives while the overdot represents a derivative 
with respect to time.  Using these expressions together with
eqs.~(\ref{ron6.6}) and (\ref{ron6.7}), we find that the stress-energy tensor
on the surface has only two nonzero components:
\be
T_2^2 = T_3^3 = -\frac{3\mun}{8\pi G} \, \delta(r-\ri) \; .
\ee
This represents precisely the tangential stress at the surface of a fluid;
that is, surface tension.  It is essentially the same as the force holding a
balloon in equilibrium \cite{Adl05}.

To summarize this part of the model, we may view the inflationary phase of our
model in PG coordinates as a small ball of de~Sitter space, which gives rise to
the fireball and later the dust ball that make up our universe, as depicted in
figure~\ref{fig11}.  If one accepts the inflationary paradigm, then the initial
singularity discussed in sections~4 and 5 may be thought of as a rough
phenomenological description of a region of de~Sitter space.

\begin{figure}[t!]
\begin{center}
\includegraphics[width=80mm]{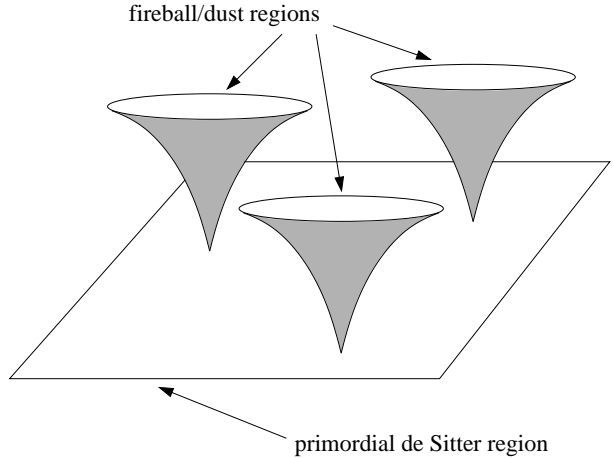}
\end{center}
\caption{\small A somewhat fanciful view of cosmogenesis, based on the
   exterior space in conformal coordinates shown in figure~\ref{fig10} but
   modified to include a de~Sitter region.}
\label{fig12}
\end{figure}

We note finally that the physical picture of cosmogenesis that was discussed
briefly in section~5 based on conformal coordinates should now be modified so
that the 3-space-filling singularity is replaced by a spacelike region of
primordial de~Sitter space, as illustrated schematically in figure~\ref{fig12}
(compare figs.~\ref{fig9} and \ref{fig10}).
Our region of the universe, and cosmic time, begin as an exploding ball of
fire which cools to a dust ball, as shown in the figure.  There is no reason
why such an initial state should not give rise to more than one such exploding
fireball, or why the fireballs produced in this way should not be able to
communicate with each other, or even to merge --- a feature that critically
distinguishes this finite-universe model from superficially similar proposals
by many other authors involving multiverses (see \cite{Sto06} and references
therein).

\section{Summary and conclusions}

We have presented what is effectively a truncated version of the standard
concordance cosmology, one that might be termed a ``truncordance model'' for
short.  This model goes over to the conventional one in the limit of a large
dust ball in which we are not too close to the edge, as should be expected.
It presents many intriguing features, which we have explored using various
coordinate systems.  We have only sketched how the model can be extended to
encompass earlier inflationary and radiation-dominated eras.  Perhaps most
satisfyingly, our model offers the possibility of experimental testability,
at least in principle, a feature which distinguishes it from other
suggestions of recent years.

\section*{Acknowledgments}

\noindent We thank R.V.~Wagoner and the members of the Gravity Probe~B theory
group for their indulgence and perceptive comments.


\begin{thebibliography}{99}
\bibitem{AO05} R.J. Adler and J.M. Overduin, {\it Gen. Rel. Grav.} {\bf 37}
   (2005) 1491
\bibitem{Chi03} L.-Y. Chiang et al., {\it Ap. J.} {\bf 590} (2003) L65
\bibitem{Lar05} D.L. Larson and B.D. Wandelt, {\it Phys. Rev. D.}, submitted;
   astro-ph/0505046
\bibitem{Toj05} R. Tojeiro et al., {\it Mon. Not. R. Astron. Soc.} {\bf 365}
   (2006) 265; astro-ph/0507096
\bibitem{Ber05} A. Bernui et al., astro-ph/0511666
\bibitem{Eri03} H.K. Eriksen et al., {\it Astrophys. J.} {\bf 605} (2004) 14;
   astro-ph/0307507
\bibitem{Eri04} H.K. Eriksen et al., {\it Astrophys. J.} {\bf 612} (2004) 64;
   astro-ph/0401276
\bibitem{Han04a} F.K. Hansen et al., {\it Astrophys. J.} {\bf 607} (2004) L67;
   astro-ph/0402396
\bibitem{Han04b} F.K. Hansen, A.J. Banday and K.M. G\'orski,
   {\it Mon. Not. R. Astron. Soc.} {\bf 354} (2004) 641; astro-ph/0404206
\bibitem{Han04c} F.K. Hansen et al.,
   {\it Mon. Not. R. Astron. Soc.} {\bf 354} (2004) 905; astro-ph/0406232
\bibitem{Cop05} C.J. Copi et al., astro-ph/0508047
\bibitem{Fre05} P.E. Freeman et al., {\it Astrophys. J.} {\bf 638} (2005) 1;
   astro-ph/0510406
\bibitem{Liu05} X. Liu and S.N. Zhang, {\it Astrophys. J.} {\bf 636} (2006) L1;
   astro-ph/0511550
\bibitem{Vie03} P. Vielva et al., {\it Astrophys. J.} {\bf 609} (2004) 22;
   astro-ph/0310273
\bibitem{Cru04} M. Cruz et al., {\it Mon. Not. R. Astron. Soc.} {\bf 356}
   (2005) 29; astro-ph/0405341
\bibitem{McE05a} J.D. McEwen et al., {\it Mon. Not. R. Astron. Soc.} {\bf 359}
   (2005) 1583; astro-ph/0406604
\bibitem{Ino06} K.T. Inoue and J. Silk, astro-ph/0602478
\bibitem{Cay05} L. Cay\'on , J. Jin and A. Treaster,
   {\it Mon. Not. R. Astron. Soc.} {\bf 362} (2005) 826; astro-ph/0507246
\bibitem{Cru06} M. Cruz et al., astro-ph/0601427
\bibitem{Tom05} K. Tomita, {\it Phys. Rev.} {\bf D72} (2005) 103506; erratum
   {\it Phys. Rev.} {\bf D73} (2006) 029901; astro-ph/0509518
\bibitem{McE05b} J.D. McEwen et al., astro-ph/0510349
\bibitem{Efs03} G. Efstathiou,
   {\it Mon. Not. R. Astron. Soc.} {\bf 343} (2003) L95 [astro-ph/0303127]
\bibitem{Uza03} J.-P. Uzan, U. Kirchner and G.F.R. Ellis,
   {\it Mon. Not. R. Astron. Soc.} {\bf 344} (2003) L65 [astro-ph/0302597]
\bibitem{Lum05} J.-P. Luminet, astro-ph/0501189
\bibitem{Pai21} P. Painlev\'e, {\it Comptes Rendus Acad. Sci.} (Paris) {\bf 173} (1921) 677
\bibitem{Gul22} A. Gullstrand, {\it Arkiv. Mat. Astron. Fys.} {\bf 16} (1922) 1
\bibitem{Kot18} F. Kottler, {\it Ann. Phys.} (Berlin) {\bf 56} (1918) 401
\bibitem{Wey19} H. Weyl, {\it Phys. Zeits.} {\bf 20} (1919) 31
\bibitem{Adl05} R.J. Adler et al., {\it Am. J. Phys.} {\bf 73} (2005) 1148
%\bibitem{Bjo01} J.D. Bjorken, {\it Phys. Rev.} {\bf D64} (2001) 085008
   %[hep-ph/0103349]
%\bibitem{Bjo03} J.D. Bjorken, {\it Phys. Rev.} {\bf D67} (2003) 043508
   %[hep-ph/0210202]
\bibitem{Sto06} W.R. Stoeger, astro-ph/0602356
\end{thebibliography}
\end{document}